\documentclass[prb,letterpaper,twocolumn,showpacs]{revtex4}

\usepackage{times,xspace}
\usepackage{amsbsy,amssymb,amsmath,bm}
\usepackage{graphicx,color,epsfig}

 \def\bbbc{{\mathchoice {\setbox0=\hbox{$\displaystyle\rm C$}\hbox{\hbox to0pt{\kern0.4\wd0\vrule height0.9\ht0\hss}\box0}} {\setbox0=\hbox{$\textstyle\rm C$}\hbox{\hbox
to0pt{\kern0.4\wd0\vrule height0.9\ht0\hss}\box0}} {\setbox0=\hbox{$\scriptstyle\rm C$}\hbox{\hbox to0pt{\kern0.4\wd0\vrule height0.9\ht0\hss}\box0}} {\setbox0=\hbox{$\scriptscriptstyle\rm C$}\hbox{\hbox to0pt{\kern0.4\wd0\vrule height0.9\ht0\hss}\box0}}}}

\newcommand{\ignore}[1]{}
\newcommand{\mComment}[1]{}
\newcommand{\gComment}[1]{}
\newcommand{\jComment}[1]{}
\newcommand{\rComment}[1]{}
\newcommand{\lComment}[1]{}

\begin{document}

\topmargin -5mm
\preprint{APS/123-QED}

\title{Ordered magnetic phases of the frustrated spin-dimer compound Ba$_3$Mn$_2$O$_8$}

\author{E. C. Samulon$^1$, Y.-J. Jo$^2$, P. Sengupta$^{3,4}$, C. D. Batista$^3$, M. Jaime$^4$, L. Balicas$^2$, I. R. Fisher$^1$}

\affiliation{$^1$Geballe Laboratory for Advanced Materials and Department of Applied Physics, Stanford University, Stanford, California 94305, USA}

\affiliation{$^2$National High Magnetic Field Laboratory, Florida State University, Tallahassee, Florida 32306, USA}

\affiliation{$^3$Theoretical Division, Los Alamos National Laboratory, Los Alamos, New Mexico 87545, USA}

\affiliation{$^4$National High Magnetic Field Laboratory, Los Alamos National Laboratory, Los Alamos, New Mexico 87545, USA}

\begin{abstract}

Ba$_3$Mn$_2$O$_8$ is a spin-dimer compound based on pairs of $S=1$, 3$d^2$, Mn$^{5+}$ ions arranged on a triangular lattice. Antiferromagnetic intradimer exchange leads to a singlet ground state in zero-field. Here we present the first results of thermodynamic measurements for single crystals probing the high-field ordered states of this material associated with closing the spin gap to the excited triplet states. Specific heat, magnetocaloric effect, and torque magnetometry measurements were performed in magnetic fields up to 32 T and temperatures down to 20 mK.  For fields above H$_{c1} \sim$ 8.7 T, these measurements reveal a single magnetic phase for $H \|c $, but two distinct phases (approximately symmetric about the center of the phase diagram) for $H \perp c$.  Analysis of the simplest possible spin Hamiltonian describing this system yields candidates for these ordered states corresponding to a simple spiral structure for $H\|c$, and to two distinct modulated phases for $H \perp c$.  Both single-ion anisotropy and geometric frustration play crucial roles in defining the phase diagram.

\end{abstract}


\maketitle

\section{\label{sec:level1}Introduction}

Antiferromagnetic exchange on a triangular lattice leads to geometric frustration - the system cannot satisfy all of the pairwise interactions simultaneously, such that the minimum energy of the system does not correspond to the minimum energy of all local interactions. The classical solution to the Heisenberg antiferromagnet on a triangular lattice with only nearest neighbor interactions is the well known 120$^{\circ}$ structure. In this case, the main effect of the frustration is simply to produce a non-colinear structure. In an early attempt to find the groundstate of the two-dimensional quantum triangular antiferromagnet, Anderson proposed the Resonating Valence Bond (RVB) spin liquid state \cite{Anderson_1973}. Subsequent theoretical work has indicated that for the simple case with only nearest neighbor  interactions the classical solution is in fact stable against quantum fluctuations, but with a much reduced ordered moment \cite{Jolicoeur_1989, Bernu_1994}. Other models corresponding to more complex lattices and interactions are still the subject of intense theoretical investigation \cite{Diep_2005}. In this paper, we experimentally examine the slightly more complex case of a triangular lattice decorated by vertical magnetic (spin) dimers, realized by the compound Ba$_3$Mn$_2$O$_8$.

Spin dimer compounds comprise pairs of strongly coupled magnetic ions. Antiferromagnetic intradimer exchange leads to a ground state that is a product of singlets, but an applied magnetic field can be used to close the spin gap to excited triplet states, resulting at low temperatures in a state characterized by long range magnetic order (LRMO) \cite{Sachdev_2004}. Close to the critical field, the effective Hamiltonian that describes the low-energy degrees of freedom of such a system can be expressed in terms of an effective spin-$\frac{1}{2}$ model, or equivalently a lattice gas model of hard-core bosons \cite{Matsubara_1956}. The exchange anisotropy in this effective model, which is sensitive to the lattice geometry and contains contributions from both the interdimer coupling and also other anisotropies present in the system, is equivalent to the balance between potential and kinetic energies in the bosonic picture.  This anisotropy plays a critical role in determining the nature of the ordered state \cite{Rice_2002}.

Spin dimer compounds offer several specific advantages over simple (non-dimerized) magnetic lattices. First, variation of an external magnetic field can be used to tune the triplet density, providing easy access to more of the quantum phase diagram, and of course the quantum critical point (QCP). The effects of quantum fluctuations at the QCP can be dramatic, especially for spins arranged on a frustrated lattice. For example, in the case of the spin dimer compound BaCuSi$_2$O$_6$, the effects of ``order from disorder" are suppressed at the QCP precisely because the size of the moment is tuned by the external field (i.e. this is an amplitude-driven QCP), and the frustration implicit in the body-centered tetragonal lattice of this material ultimately leads to a form of dimensional reduction \cite{Sebastian_2006a, Batista_2007}. Second, the interdimer exchange energy ``protects" the system against symmetric anisotropies, effectively suppressing the influence of interactions such as dipolar coupling and, for the case in which the kinetic energy of the bosons dominates, enabling a realization of a Bose Einstein condensate (BEC) in a temperature range which is even comparable to the anisotropy energy \cite{Sebastian_2006b}. (This is why one specifically looks for such an effect in a spin dimer compound and not a simple antiferromagnet.) And third, spin dimer compounds provide a means to ``engineer" large exchange anisotropies, providing access to lattice gas models that would otherwise require unphysical parameters for a simple antiferromagnet. Specifically, although individual exchange couplings in such materials may be nearly isotropic ($J_z \sim J_{xy}$), in the strong coupling limit (intradimer exchange $J \gg$ interdimer exchange $J'$) the effective Hamiltonian derived from perturbation theory and acting on the singlet and triplet states can have strongly anisotropic effective exchange couplings ($J^{eff}_z \neq J^{eff}_{xy}$).   As an example, in some circumstances a large uniaxial anisotropy ($J^{eff}_z \gg J^{eff}_{xy}$) would provide favorable conditions for realization of a spin supersolid phase \cite{Sengupta_2007}.

In this paper we examine the high field behavior of the novel $S = 1$ triangular spin dimer compound Ba$_3$Mn$_2$O$_8$. The interest in this specific compound stems from the possibility to explore the interplay between geometric frustration (which typically favors uniform spiral states), and single ion anisotropy (favoring specific orientations of the moments) in the context of a spin dimer compound, in which case we can explore the entire quantum phase diagram, ultimately including the quantum critical behavior. We determine the magnetic phase diagram via heat capacity, magnetocaloric effect and torque magnetization measurements of single crystals, finding a single magnetic phase for fields aligned parallel to the crystalline $c$-axis, but two magnetic phases for fields oriented away from this direction. Comparison of these results with a low energy effective model indicate that the competing effects of interdimer coupling and single ion anisotropy on the triangular lattice stabilize complex magnetic structures characterized by multiple independent order parameters, including spontaneous bond order.

\begin{figure}
\includegraphics[width=8cm]{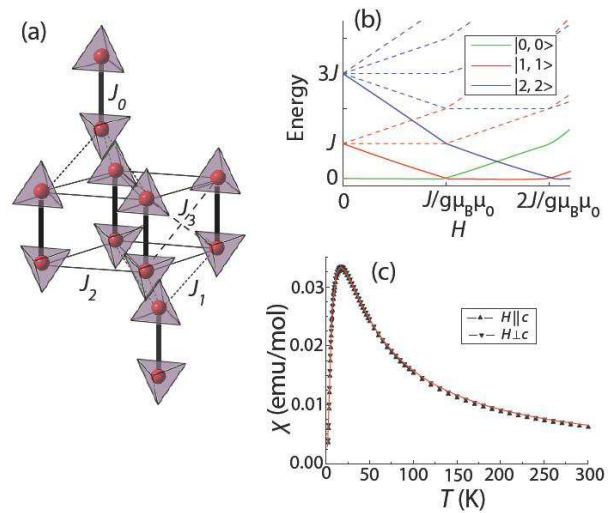}
\caption{(Color online) (a) Schematic diagram showing arrangement of MnO$_4$ tetrahedra in R$\bar{3}$m structure of Ba$_3$Mn$_2$O$_8$. Intradimer ($J_0$), out-of-plane interdimer ($J_1$), in-plane direct interdimer ($J_2$) and in-plane crossed interdimer ($J_3$) exchange bonds are drawn as thick black, dotted black, thin black and dashed black lines, respectively\cite{Uchida_2002}. Only two of the in-plane crossed interdimer exchange bonds ($J_3$) were drawn for clarity.  (b) Energy spectrum as a function of field for an isolated dimer composed of $S$=1 Mn$^{5+}$ ions with antiferromagnetic exchange $J$.  (c) Susceptibility of Ba$_3$Mn$_2$O$_8$ as a function of temperature for $\mu_0H$ = 0.1T applied parallel (up triangle) and perpendicular (down triangle) to the $c$-axis. Mol refers to one dimer unit. Red line shows fit to dimer model as described in main text.}
\label{XtalSuscept}
\end{figure}

Ba$_3$Mn$_2$O$_8$ crystallizes in the rhombohedral R$\bar{3}$m structure, and is comprised of pairs of Mn$^{5+}$ 3$d^2$ $S=1$ ions arranged vertically on hexagonal layers (see fig. \ref{XtalSuscept}(a))\cite{Weller_1999}. Each Mn ion is coordinated by distorted oxygen tetrahedra, which results in an orbitally non-degenerate ground state. Successive layers are stacked following an `$ABC$' sequence, such that the dimer units on adjacent planes are positioned in the center of the triangular plaquets of the layers above and below. Pairs of ions on each dimer are coupled antiferromagnetically, resulting in a singlet ground state (fig. \ref{XtalSuscept}(b)). Initial powder inelastic neutron scattering (INS) indicated an intradimer exchange energy $J_0 \sim$ 1.61(3) meV \cite{Stone_2008}. The same INS measurements revealed a spin gap of 1.05 meV and a hierarchy of additional exchange energies, in which interdimer coupling between Mn ions in the same plane is characterized by $J_2 - J_3 = 0.112^{+0.015}_{-0.003}$ meV and interdimer coupling between Mn ions residing on adjacent planes is characterized by $J_1 = -0.062^{+0.007}_{-0.066}$ meV.  Preliminary single crystal INS measurements further refined these values to $J_0=1.65$meV, $J_2-J_3=0.109$meV and $J_1=-0.120$meV \cite{Stone_2008b} (Here we preserve the labelling of exchange energies initially suggested by Uchida \textit{et al.} \cite{Uchida_2002}).  Ba$_3$Mn$_2$O$_8$ can therefore be described as a quasi-2D material in which planes of vertical dimers arranged on triangular layers interact weakly in the perpendicular direction \cite{Interaction}. Electron paramagnetic resonance (EPR) experiments in the diluted compound Ba$_3$(V$_{1-x}$Mn$_x$)$_2$O$_8$ (where the V$^{5+}$ 3$d^0$ ion carries no moment) reveal a single ion uniaxial anisotropy characterized by $D$ = 5.81 GHz, corresponding to 0.024 meV \cite{Whitmore_1993}. Similar measurements for the pure compound Ba$_3$Mn$_2$O$_8$ indicate a zero field splitting of the triplet characterized by $|D|$ = 0.032 meV \cite{Hill_2007}, the difference being due to the presence of additional symmetric anisotropies in the concentrated lattice, in particular dipolar coupling between the two ions on each dimer. Values of the $g$-tensor were revealed by EPR measurements of the diluted compound to be $g_{cc} = 1.96$ and $g_{aa} = g_{bb} =1.97$ \cite{Whitmore_1993}, and are confirmed by EPR measurements in the concentrated lattice \cite{Hill_2007}.

The spin gap in Ba$_3$Mn$_2$O$_8$ can be closed by an applied field $H_{c1} \sim$ 9 T, and measurements of powder samples have revealed a magnetization which rises approximately linearly with field from $H_{c1}$  until the eventual triplet saturation field is reached at $H_{c2} \sim$ 26 T \cite{Uchida_2002}. Heat capacity measurements of polycrystalline samples revealed tantalizing evidence for two phase transitions \cite{Tsujii_2005}, but to date no single crystal samples have been available to definitively determine the phase diagram, nor probe the ordered states. A second spin gap to excited quintuplet states can be closed by increasing the field beyond $H_{c2}$, leading to an additional increase in magnetization above approximately 30 T \cite{Uchida_2002}.  In this paper we investigate solely the nature of the long range magnetic order which results from closing the spin gap to the triplet states.

\section{\label{sec:level2}Experimental Methods}

Single crystals of Ba$_3$Mn$_2$O$_8$ were grown by a slow cooling flux method. Polycrystalline precursor was synthesized by mixing MnO and BaCO$_3$ reagents and sintering at temperatures up to 1000$^{\circ}$ C for 100 hours with intermediate regrindings according to the previously published method\cite{Uchida_2002}. To grow crystals of Ba$_3$Mn$_2$O$_8$ from solution requires a strongly oxidizing flux - we have found anhydrous NaOH to work especially well. Polycrystalline material was mixed with NaOH in a molar ratio of 1 to 19 and placed in a 20 cc alumina crucible lightly sealed with a cap. The mixture was heated to 700$^{\circ}$ C over the course of 24 hours, allowed to dwell for 5 hours, and then slowly cooled to 300$^{\circ}$ C at which temperature the furnace was turned off. The flux can be removed by repeated washes in water. Crystals grown by this method form as hexagonal tablets, and have a mass of up to 100 mg. From initial X-ray diffraction measurements they have a small mosaic spread of less than 0.10$^{\circ}$, and minimal magnetic impurities (see characterization in following section).

Low-field susceptibility measurements were performed using a commercial Quantum Design MPMS XL SQUID magnetometer for fields of 1000 Oe applied both parallel and perpendicular to the $c$-axis.

Heat capacity ($C_p$) data were collected down to 0.35 K using a thermal relaxation time technique, both in a Quantum Design Physical Properties Measurement System (PPMS) for fields up to 14 T applied parallel to the $a$ and $c$ axes, and at the National High Magnetic Field Laboratory (NHMFL) using a home-built calorimeter in a resistive magnet for fields up to 32 T applied parallel to the $a$-axis. Results were repeated for several crystals. Magnetocaloric effect (MCE) scans were also performed in the resistive magnet at the NHMFL using the same home-built calorimeter used for heat capacity measurements. Data were taken for sweep rates of 2 T/min and 5 T/min, for both increasing and decreasing fields, and for fields oriented both parallel and perpendicular to the $c$-axis.

Cantilever torque measurements were performed at the NHMFL for fields up to 18 T using a superconducting magnet. The sample was mounted on a cantilever forming one plate of a capacitor and oriented such that the magnetic field was slightly misaligned with one of the principle axes. As a consequence of the misalignment and the intrinsic $g$-anisotropy, the resulting magnetization was not parallel to the applied field, resulting in a finite torque. Although the $g$-anisotropy of Ba$_3$Mn$_2$O$_8$ is relatively weak \cite{Whitmore_1993}, the resulting torque is nevertheless measurable and can be used to provide a sensitive probe of changes in the magnetization close to $H_{c1}$, and in particular at the phase transition(s). The cantilever was mounted on the cold finger of a dilution refrigerator, and field sweeps were performed for temperatures from 20 to 820 mK.

\section{\label{sec:level3}Results}

The low field susceptibility of Ba$_3$Mn$_2$O$_8$ is shown in fig. \ref{XtalSuscept}(c) for fields aligned parallel and perpendicular to the $c$-axis. There is negligible anisotropy, consistent with the small $g$-anisotropy determined from EPR measurements \cite{Whitmore_1993, Hill_2007}. As previously described for polycrystalline samples \cite{Uchida_2002}, the temperature dependence of the susceptibility can be well fit to an isolated $S$=1 dimer model:
\begin{equation}
\chi_d = \frac{2N\beta g^2 \mu_B^2 \left(1+5e^{-2\beta J}\right)}{3+e^{\beta J} + 5e^{-2 \beta J}},
\end{equation}
if one includes a meanfield correction to account for exchange with neighboring dimers of the form $\chi = \frac{\chi_d}{1+\lambda \chi_d}$ where $\lambda = 3\left[J_1+2\left(J_2 + J_3\right) \right] / \left( Ng^2\mu_B^2 \right)$ and $\beta = 1/k_BT$. The fit also includes terms to account for a temperature-independent background ($\chi_0$) and a small concentration of paramagnetic impurities ($C/T$). These fits result in estimates of $J_0$ = 1.44$\pm$0.01 meV and $g_c \sim g_a \sim 2.01\pm$0.03, which are close to the values obtained from INS \cite{Stone_2008} and EPR \cite{Hill_2007} experiments, and also to similar susceptibility fits for polycrystalline samples \cite{Uchida_2002}. The fit is rather insensitive to the precise value of $\lambda$ (which is why this is a poor method to estimate interdimer exchange coefficients), but nevertheless returns a best value of $\lambda= 5.0 \pm$ 0.3 mol/emu which is remarkably close to the calculated value of 6.6 mol/emu based on estimates of the exchange constants 3$J_1$ + 6$\left(J_2 + J_3\right)$ = 0.83 meV obtained from single crystal INS measurements \cite{Stone_2008}. The T-independent term has a value $\chi_0$= 2x10$^{-4}$ emu/mol, and the impurity Curie term corresponds to just 0.4$\%$ unpaired Mn$^{5+}$ ions.

\begin{figure}
\includegraphics[width=8cm]{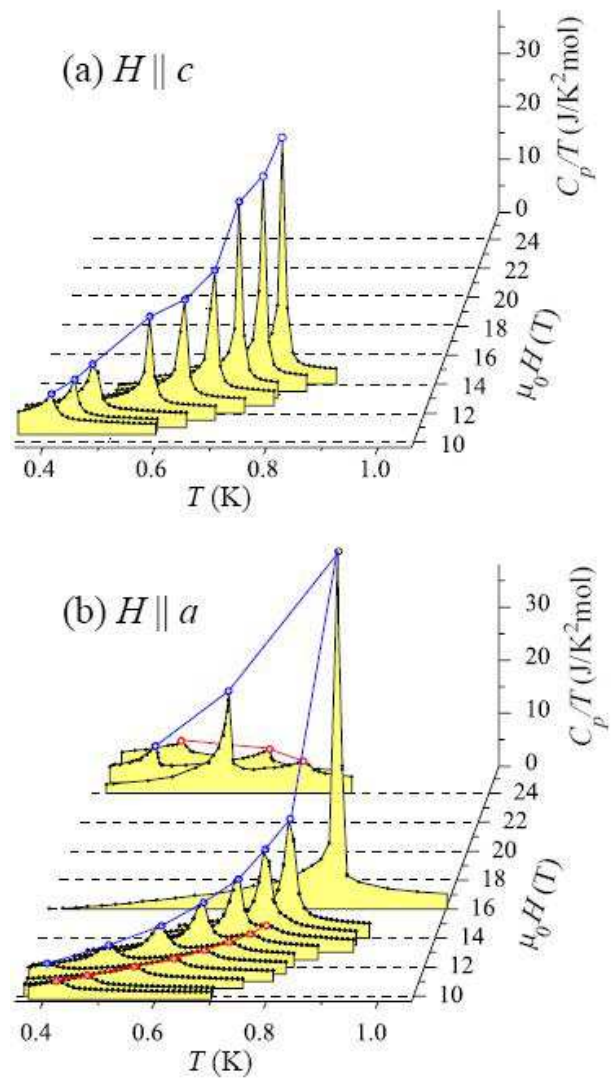}
\caption{(Color online) Heat capacity (shown as $C_p$/$T$) as a function of temperature for fields applied (a) parallel and (b) perpendicular to the c-axis. Blue (red) lines indicate the transition into Phase I (Phase II) from higher temperatures as determined by peaks in $C_p$/$T$ as a function of field.}
\label{Cp}
\end{figure}

\begin{figure}
\includegraphics[width=8cm]{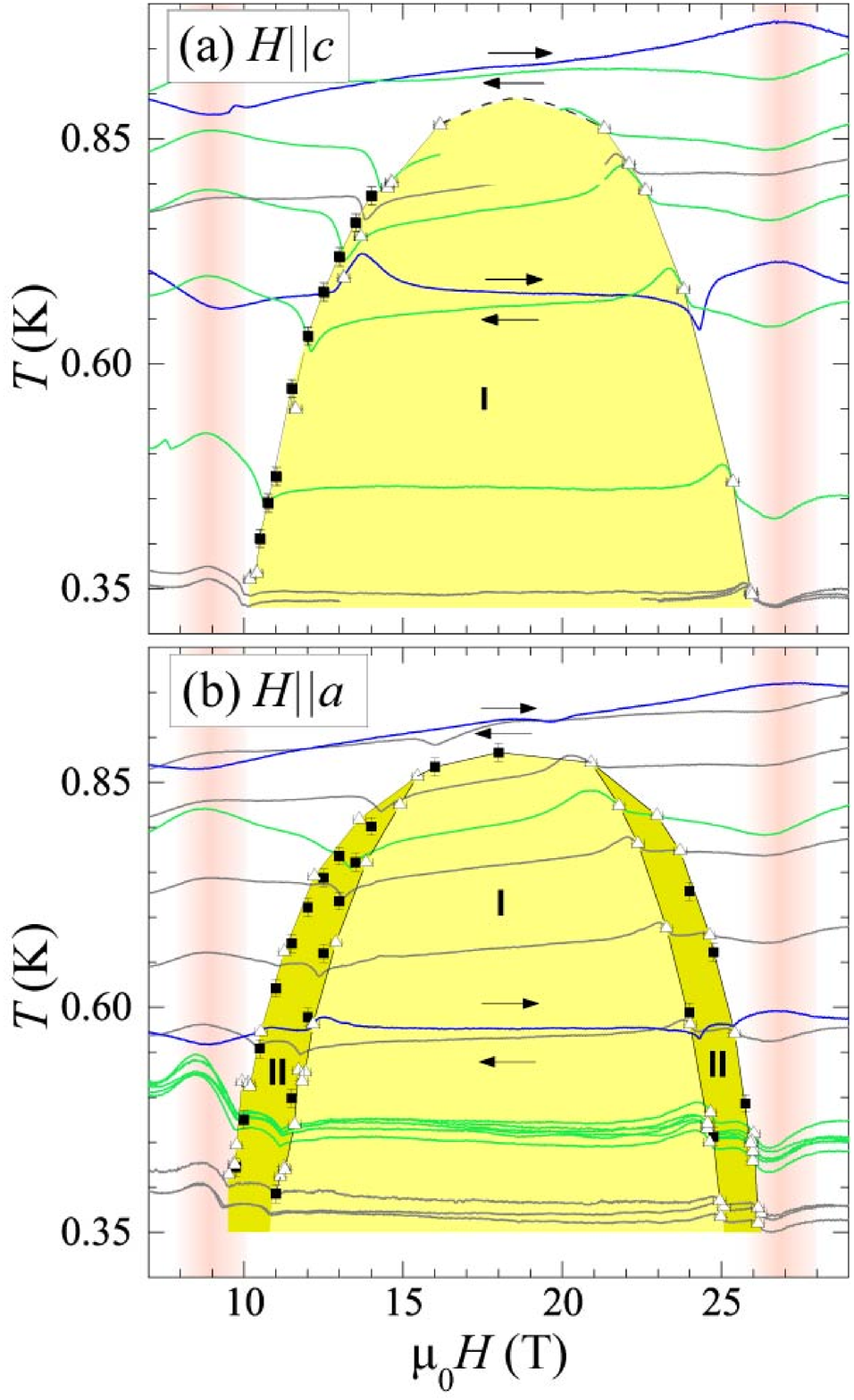}
\caption{(Color online) Phase diagram obtained from heat capacity (solid squares) and MCE (open triangles) measurements for fields applied (a) parallel and (b) perpendicular to the c-axis.  MCE traces are shown in gray (green) for decreasing fields for sweep rates of 2 (5) T/min. Representative data for increasing fields are designated by arrows and shown in blue for sweep rates of 5 T/min in panel (a) and 2 T/min in panel (b).  Labels indicate Phases (I) and (II), as described in the main text. Dashed line in panel (a) indicates the anticipated phase boundary for fields between 16 and 21 T based on similar data in panel (b). Shaded vertical bands are guides to the eye to draw attention to the broad features observed in MCE measurements centered at 8.8T and 26.5T.}
\label{PhaseB}
\end{figure}

Heat capacity data are shown in fig. \ref{Cp} for temperatures down to 0.35 K. For fields greater than $H_{c1}$ applied parallel to the $c$-axis (fig. \ref{Cp}(a)) there is just one phase transition in this temperature range. We label this ordered state Phase I to distinguish from a distinctly different phase observed for fields parallel to the $a$-axis. The data show a lambda-like transition, suggestive of 3DXY ordering, with an integrated entropy which increases as the field is increased from $H_{c1}$ up to the maximum field for which data were taken for this orientation, which was 14 T. These data points are included as solid square symbols in the phase diagram shown in fig. \ref{PhaseB}(a).

\begin{figure}
\includegraphics[width=8cm]{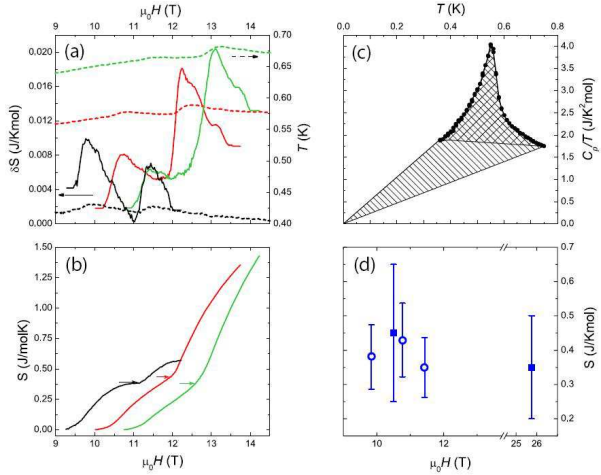}
\caption{(Color online) Entropy estimates for the Phase II for fields oriented parallel to the $a$-axis.  (a) Three representative MCE measurements (dashed lines, right axis) and the corresponding change in entropy, $\delta S_i$, (solid lines with the same color, left axis). (b) Total entropy associated with phase transitions seen in MCE measurements, calculated as described in the main text. Horizontal arrows indicate the entropy associated with the lower transition. (c) Upper and lower bounds of the integrated entropy associated with the phase transition observed in heat capacity at 10.5T. (d) Entropy on entering Phase II from MCE (open circles) and $C_p$ (solid squares).}
\label{Entropy}
\end{figure}

For fields oriented parallel to the $a$-axis, heat capacity data were taken up to $H_{c2} \sim$ 26 T (fig. \ref{Cp}(b)). These data show a remarkable sequence of phase transitions at low temperature with an unusual division of entropy. For fields between 9 and 11 T just one transition is observed above 0.35 K; between 11 and 13 T two distinct transitions are clearly resolved; for intermediate fields, only a single transition; between 24 and 25 T two transitions are again observed; and finally close to the triplet saturation field only one transition is observed. Anomalies in the heat capacity marking these phase transitions are joined by red and blue lines in fig. \ref{Cp}(b), and $T_c$ values included in the phase diagram shown in fig. \ref{PhaseB}(b) as solid symbols. The heat capacity anomaly for phase transitions joined by the blue line in fig. \ref{Cp}(b) are lambda-like, similar to those observed for fields parallel to the $c$-axis, and accordingly we label this state Phase I. The integrated entropy associated with these transitions first rises with field, and then after $H$ is increased beyond the midpoint of the phase diagram, reduces in magnitude again. In contrast, the anomaly associated with the phase transitions connected by the red lines in fig. \ref{Cp}(b) are less divergent, and although the data do not permit a critical scaling analysis, nevertheless are more suggestive of an Ising transition. We refer to this state as Phase II. The rise in $C_p/T$ associated with this phase transition does not appear to vary with field within the uncertainty, indicating that the change in entropy is only weakly dependent on the applied field. Estimates of the integrated entropy are difficult due to the close proximity of the second phase transition, and also due to the large background magnetic contribution to the heat capacity associated with the other gapped states. However, a crude estimate of this entropy was obtained for several fields for which the only resolvable transition is between the disordered phase and Phase II (plotted in fig. \ref{Entropy}(c) for $H= 10.5$ T).  Upper and lower bounds for the entropy were determined by assuming a minimum and maximum possible background, shown as blue and red lines respectively in fig. \ref{Entropy}(c), yielding an average of 0.45 $\pm$ 0.20 J/molK. Within the uncertainty, this value appears to be symmetric for fields above and below the midpoint of $H_{c1}$ and $H_{c2}$ as shown in fig. \ref{Entropy}(d), and for this reason we also label the ordered state on the right hand side of Phase I in fig. \ref{PhaseB}(b) as Phase II.

Additional insight to these phase transitions is provided by MCE measurements. Phase transitions are evident from a sharp increase (decrease) in the temperature of the sample on entering (leaving) the ordered state. In practice, points on the left (right) hand side of the phase diagram (open symbols in fig. \ref{PhaseB}) were determined from a sharp peak (trough) in the first derivative of the temperature with respect to field taken on up (down) field sweeps, each corresponding to the case of entering the ordered state. These data are in close agreement with heat capacity measurements, with small differences being ascribed to differences in sample alignment and, where two different calorimeters were used, thermometry. A slight assymmetry in the magnitude of the change in temperature between up and down sweeps provides evidence that phase boundaries between the disordered state and Phase I, between the disordered state and Phase II, and between ordered Phases I and II are all weakly first order in this temperature range.

The total change in entropy associated with a phase transition can be calculated from MCE measurements by summing the increase in entropy of the system plus the entropy lost as heat to the bath from the sample stage:
\begin{equation}
\delta S_i = -\frac{C\left(T_{i+1} - T_i\right) + \kappa \left(T_i - T_{base}\right)}{T_i},
\label{MCE_equation}
\end{equation}
where $i$ labels successive temperature points taken as the field is swept (typically 40 evenly spaced points per Tesla) and $\kappa$ is the thermal conductivity of the thermal link in the calorimeter. A linear interpolation for $\kappa$ was calculated as a function of temperature and field for the calorimeter, and values of the heat capacity were taken from measurements performed in the PPMS calorimeter. Figure \ref{Entropy}(a) shows three representative MCE data sets for fields oriented perpendicular to the $c$-axis (dashed line, right axis), and the associated change in entropy $\delta S_i$ between successive data points (solid curves, left axis). All three data sets were taken for increasing fields and for the same sweep rate of 2 T/min. As can be seen, $\delta S_i$ shows two successive peaks as a function of field, which correspond to the two phase transitions. A practical measure of the change in entropy associated with each phase transition is therefore provided by the integrated entropy up to the minimum in $\delta S_i$, which is shown in fig. \ref{Entropy}(b). For the lowest temperature data set (black curves) the two transitions are well separated and the total entropy associated with the first transition exhibits a clear plateau. For the higher temperature sweeps (red and green curves) the two transitions are slightly closer in field, and the total entropy exhibits more of a kink than a plateau. Nevertheless, these data allow an estimate of the integrated entropy associated with each transition, which are plotted in fig. \ref{Entropy}(d) for these and some additional intermediate temperature sweeps. Within the uncertainty of this analysis, the change in entropy associated with entering Phase II from the disordered state is essentially independent of temperature, with an absolute value that agrees remarkably well with the value extracted from heat capacity measurements (square symbol in fig. \ref{Entropy}(d)). In contrast, the change in entropy associated with entering Phase I depends strongly on temperature, consistent with inspection of the heat capacity data shown in fig. \ref{Cp}(b).

The MCE experiments did not reveal any evidence for additional phase transitions other than those indicated in fig. \ref{PhaseB}. However, broad features were observed centered at $H_{c1}$ and $H_{c2}$ for both field orientations (vertical shading in fig. \ref{PhaseB}). This effect is associated with the rapid change in magnetization with temperature at these fields. Specifically, Maxwell's relation $\frac{\partial M}{\partial T}|_{H,P}=\frac{\partial S}{\partial H}|_{T,P}$ implies a large change in the entropy at $H_{c1}$ and the triplet saturation field for field sweeps performed at constant temperature, or conversely, a large change in the temperature of the sample in a constant entropy experiment. The effect reverses sign for increasing/decreasing fields, and for entering/leaving the field region between $H_{c1}$ and the triplet saturation field, and becomes rapidly smeared out at higher temperature.

The phase diagram obtained from heat capacity and MCE measurements (fig. \ref{PhaseB}) is almost but not completely symmetric about its midpoint. For fields oriented parallel to the $a$-axis (fig. \ref{PhaseB}(b)), Phase II is slightly narrower in field on the right hand side of the phase diagram relative to the left. The absence of strict particle-hole symmetry is consistent with the large bandwidth relative to the spin gap observed in powder INS \cite{Stone_2008}.

\begin{figure}
\includegraphics[width=8cm]{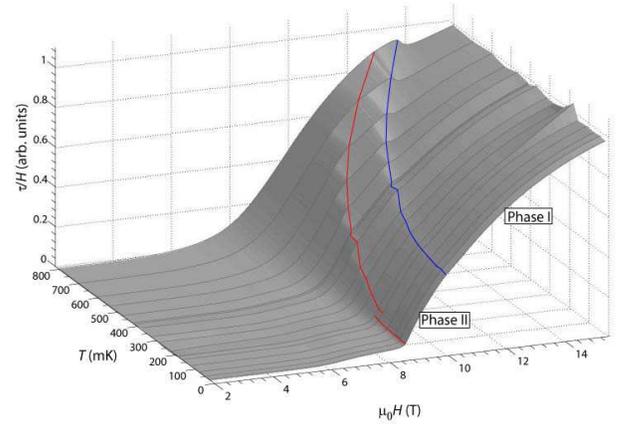}
\caption{(Color online) Field and temperature dependence of the magnetic torque (scaled by the field strength) for fields applied close to perpendicular to the $c$ axis. A linear interpolation scheme between field sweeps taken at 20, 45, 75, 114, 160, 183, 237, 280, 318, 372, 412, 417, 470, 535, 600, 696 and 800 mK (dark lines) has been used to generate the 3D surface.  Red and blue lines indicate phase transitions determined as described in the main text.  Labels indicate Phases I and II.}
\label{3DTorq}
\end{figure}

\begin{figure}
\includegraphics[width=8cm]{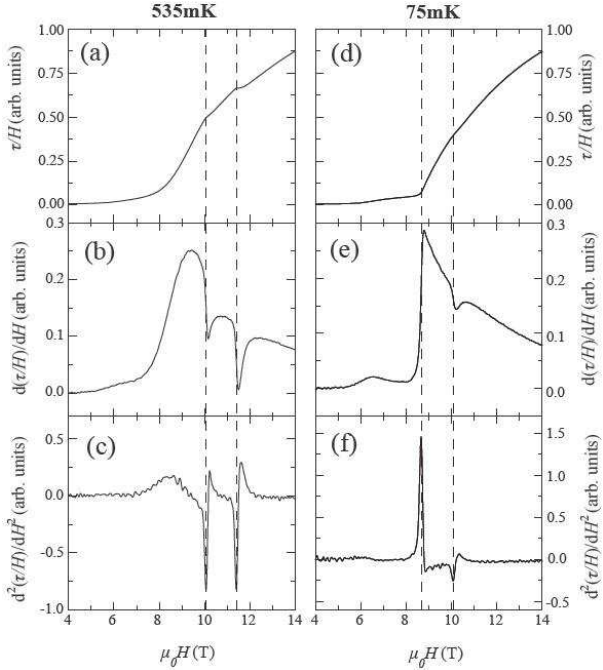}
\caption{Representative cantilever torque measurements for fields applied close to perpendicular to the $c$-axis, expressed as $\tau/H$ and its first two derivatives for temperatures of 535 mK (a-c) and 75 mK (d-f). Vertical dashed lines mark phase transitions as described in text}
\label{TwoTorq}
\end{figure}

\begin{figure}
\includegraphics[width=8cm]{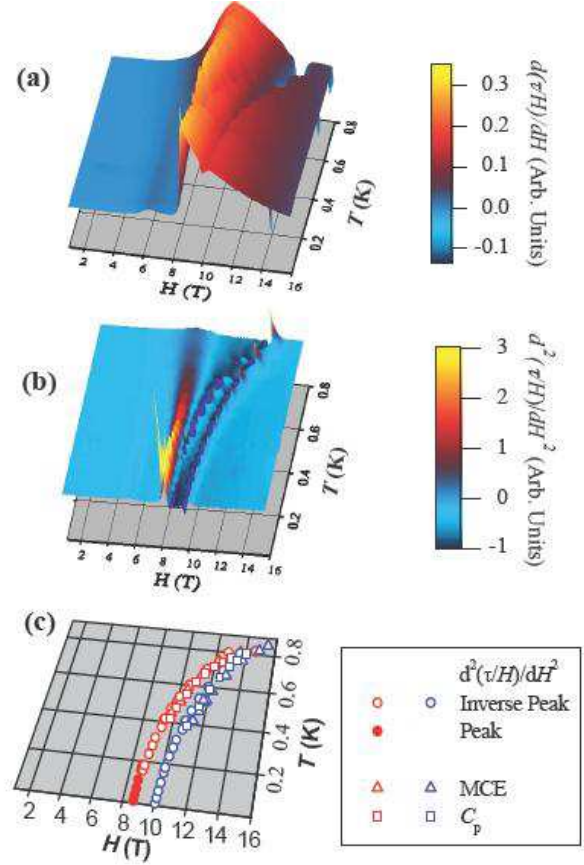}
\caption{(Color online) (a) First and (b) second derivatives with respect to field of torque divided by field for fields close to perpendicular to the c axis.  (c) Phase diagram extracted from MCE (triangles), C$_p$ (squares), and torque (circles).  Open symbols are used to signify a weakly 1$^{st}$ order transition.  Closed symbols signify a second order transition. Red (blue) symbols indicate the lower (upper) field transition.}
\label{TorqDer}
\end{figure}

Further information about the magnetically ordered phases can be gained from torque magnetometry. Raw data taken for fields nearly perpendicular to the $c$-axis, scaled by the magnetic field strength, are shown as a 3D surface in fig. \ref{3DTorq}. The sharp increase in torque at 8.5 T for the lowest temperatures corresponds to the spin gap closing as the minimum of the the $S_z$=1 triplet band crosses the singlet leading to a finite magnetization. As temperature is increased, thermal effects smear this rapid increase in torque, and field derivatives of $\tau/H$ at $H_{c1}$ rapidly broaden with temperature (figures \ref{TwoTorq} and \ref{TorqDer}). Superimposed on top of the broad rise in torque with field, the two phase transitions seen in heat capacity and MCE experiments are clearly visible as breaks in the slope of $\tau/H$, and points on the phase boundary can be extracted from field derivatives as described below.

Representative torque data and their derivatives are shown in fig. \ref{TwoTorq}(a-c) and (d-f) for temperatures of 535 and 75 mK respectively. Specifically, the data in panels (a-c) are representative of field scans for temperatures above 160 mK, whereas panels (d-f) are representative of field scans below 160 mK.  Considering first the 535 mK data, both phase transitions are characterized by a decrease in the slope of $\tau/H$ above each critical field (fig. \ref{TwoTorq}(a)). The first derivative of $\tau/H$ (fig. \ref{TwoTorq}(b)) more clearly reveals this effect, exhibiting a relatively broad minimum close to each transition. Inspection of fig. \ref{TwoTorq}(a) also reveals a distinct downwards ``hook" in $\tau/H$ at both critical fields, being more prominant for the second transition. Extrapolation of $\tau/H$ from above and below the two transitions reveals a discontinuous (downwards) change at the critical field, indictive of a first order transition. In this case, an empirical estimate of the critical field can be found in the sharp inverse peak in the second derivative (fig. \ref{TwoTorq}(c)) which effectively marks the onset of the change in magnetization (dashed line in fig. \ref{TwoTorq}). From simple energetic reasons the magnetization should always rise as the field is increased, therefore these features indicate a change in the anisotropy of the ordered states at the critical fields. The absence of a delta-like transition in the heat capacity indicates that both transitions are only weakly first order.

Below 160 mK, typified by data shown in figures \ref{TwoTorq}(d-f), the upper transition is still marked by a distinct decrease in the slope of $\tau/H$ as the field is increased, and features in the two field derivatives at this transition are similar to those seen at higher temperatures in fig.s \ref{TwoTorq}(a-c). The data do not exhibit as clear of a ``hook", but the smooth evolution of the raw data and both derivatives from the behavior at high temperatures implies that this transition is still weakly first order. However, for temperatures below 160 mK the lower field transition is qualitatively different to the upper transition. Specifically, the torque exhibits a rapid increase in slope (panel (d)), leading to an abrupt step in the first derivative (panel (e)) and a sharp peak in the second derivative (panel(f)). This behavior is consistent with a second order phase transition, in which case consideration of the critical exponents associated with this thermal phase transition implies that the sharp positive peak in the second derivative marks the critical field (vertical dashed line) \cite{Universality}.

Figures \ref{TorqDer} (a) and (b) show 3D surface plots of first and second field derivatives of $\tau/H$ respectively for all temperatures measured, demonstrating the smooth evolution of the various features described above. Points on the resulting phase diagram, extracted as described above, are shown in fig. \ref{TorqDer}(c) and are in agreement with points taken from MCE and $C_p$ measurements up to the inherent angular misaligment in this torque measurement. For temperatures near 160mK (i.e. in the temperature range at which the phase transition changes from 2$^{nd}$ to 1$^{st}$ order) there is some ambiguity as to which feature marks the lower field transition, and in these cases values obtained from both criteria are plotted. Unfortunately there are insufficient data points below 160 mK to extract a meaningful critical scaling exponent for the phase boundary approaching the QCP.

Careful inspection of fig. \ref{3DTorq} reveals a small rise in the torque at approximately 6 T for the lowest temperatures. No sharp features are associated with this rise, indicating the absence of any ordering transitions, and it is rapidly smeared out with increasing temperature (fig. \ref{TorqDer}(a)). The observation of a finite magnetization for fields below $H_{c1}$ indicates the presence of terms in the spin Hamiltonian that mix singlet and triplet states, such as antisymmetric Dzyaloshinskii-Moriya (DM) interactions. The midpoint of the dimer unit in Ba$_3$Mn$_2$O$_8$ is a center of inversion symmetry, ruling out the presence of an intra-dimer DM term. However, DM interactions are still possible on the inter-dimer bonds, both within planes and between adjacent planes. The absence of any torque at low fields indicates that such terms are relatively weak. Further EPR measurements should clarify the origin of this feature.

\section{\label{sec:level4}Discussion}

In the absence of measurements that directly determine the magnetic structure of the ordered phases of Ba$_3$Mn$_2$O$_8$, we resort to an analysis of the spin Hamiltonian which describes the system:
\begin{eqnarray}
\mathcal{H} & = & \sum_{i, j, \mu, \nu} \frac{J_{i \mu j \nu}}{2}\textbf{S}_{i\mu}\cdot \textbf{S}_{j\nu}
+ D\sum_{i, \mu}\left(S^{\eta}_{i \mu}\right)^2
\nonumber\\
&& - g_{\alpha \alpha} \mu_B H \sum_{i \mu} S^{z}_{i \mu}
\label{RealHam}
\end{eqnarray}
Here $i, j$ designate the coordinates of the dimers while $\mu, \nu = {1, 2}$ denote each of the two spins on a given dimer. The intra-dimer exchange interaction is $J_0$ = $J_{i1i2} = J_{i2i1}$. The nearest-neighbor (NN) inter-dimer exchange interaction on the same layer is $J_2 = J_{i \mu j \mu}$ and $J_3 = J_{i \mu j \nu}$ with $\mu \neq \nu $, where the dimers $i, j$ belong to the same layer. Finally, the NN interlayer interaction is given by $J_1 = J_{i2j1}$, where $i, j$ denote the position of NN dimers on adjacent layers. The quantization axis ($z$-axis) is always along the field direction. The components $\alpha = a, b, c$ and $\eta = {x, y, z}$ depend on the field orientation; i.e. $\alpha = c$, $\eta$ = $z$ for $H \| c$,  and $\alpha = a, b$, $\eta = x$ for $H \perp c$. We use values for the exchange parameters determined by inelastic neutron scattering \cite{Stone_2008, Stone_2008b}, and values for $D$ and the $g$-tensor determined by EPR \cite{Whitmore_1993, Hill_2007}, as described in the Introduction. The midpoint of the magnetic dimer is a center of inversion symmetry, so we need not  consider an intradimer Dzyaloshinskii-Moriya (DM) interaction. The effect of intradimer dipolar coupling is implicitly contained within the anisotropy term $D\left(S^{\eta}\right)^2$ together with the single ion anisotropy. For this initial analysis we assume that interdimer dipolar and DM interactions are negligible, although some subtle features of the torque magnetization may require inclusion of these terms for a more complete description.

Since $J_0 \gg J_1, J_2, J_3$, we can include the inter-dimer terms perturbatively relative to the intradimer Heisenberg term. For this purpose, we keep only the low energy singlet, $|00\rangle$, and triplet $|11\rangle$ states of the single dimer problem for $H \sim H_{c1}$.  We use a pseudospin $s = \frac{1}{2}$ to represent the low-energy singlet and triplet states: $|00\rangle \!\! \rightarrow  \!\! | \!\! \downarrow\rangle$ and $|11\rangle \!\! \rightarrow \!\! | \!\! \uparrow \rangle$. The low-energy effective Hamiltonian to first order in $J_1, J_2, J_3$ and $D$, $\tilde{\mathcal{H}}$, results from projecting $\mathcal{H}$ into the low-energy subspace generated by the states $|00\rangle$ and $|11\rangle$:
\begin{eqnarray}
\tilde{\mathcal{H}} & = & \frac{4J_1}{3}\sum_{l \langle \langle i, j \rangle \rangle} \left[\textbf{s}_{il} \cdot \textbf{s}_{jl+1}-\frac{13}{16} s_{il}^z s_{jl+1}^z \right]
\nonumber\\
&& + \sum_{l \langle i, j \rangle} \left[ \frac{8\left(J_2-J_3\right)}{3} \left(s_{il}^x s_{jl}^x + s_{il}^y s_{jl}^y \right) +\frac{\left(J_2+J_3\right)}{2} s_{il}^z s_{jl}^z \right]
\nonumber\\
&& + J_1 a(\eta) \sum_{l \langle \langle i, j \rangle \rangle} \left(s_{il}^x s_{jl+1}^x - s_{il}^y s_{jl+1}^y \right)
\nonumber\\
&& + 2\left(J_2-J_3\right) a(\eta) \sum_{l \langle i, j \rangle} \left(s_{il}^x s_{jl}^x - s_{il}^y s_{jl}^y \right)
\nonumber\\
&&   - B \sum_{l, i} s_{il}^z
\label{EffHam}
\end{eqnarray}
where each dimer has been replaced by an effective site,
$B = g_{\alpha \alpha} \mu_B H - J_0 - 3\left(J_2+J_3\right)/2 - 3J_1/4 -  D \delta_{\eta z}/6$, $l$ is the layer index, $\langle i,j \rangle$ indicates that the sites $i$ and $j$ are nearest neighbors (NN) on the same layer and $\langle \langle i, j \rangle \rangle$ denotes NN on adjacent layers. In addition we have $a(z) = 0$ (the model is U(1) invariant for $H \| c$) and $a(x) = - 8D/3J_0$ (see Appendix A). Although the exchange anisotropy terms are of higher order ( ${\cal O}(J_i D/J_0)$ with $i=1,2$) than the rest of the terms, we need to include them to give account of the observed differences between the $H\parallel c$ and  $H \perp c$ cases.

To understand this effective Hamiltonian, and some properties of its ground states, it is simplest to consider the two experimentally observed cases - fields applied parallel and perpendicular to the $c$-axis. In each case we first consider the 2D lattice, effectively setting $J_1 = 0$ (i.e. no interplane coupling), and then we consider the full 3D case. Finally we compare how the interlayer frustration is relieved in Ba$_3$Mn$_2$O$_8$ while the interlayer frustration is preserved in the similar 2D spin dimer compound BaCuSi$_2$O$_6$.

\subsection{Fields parallel to $c$}

\begin{figure}
\includegraphics[width=8cm]{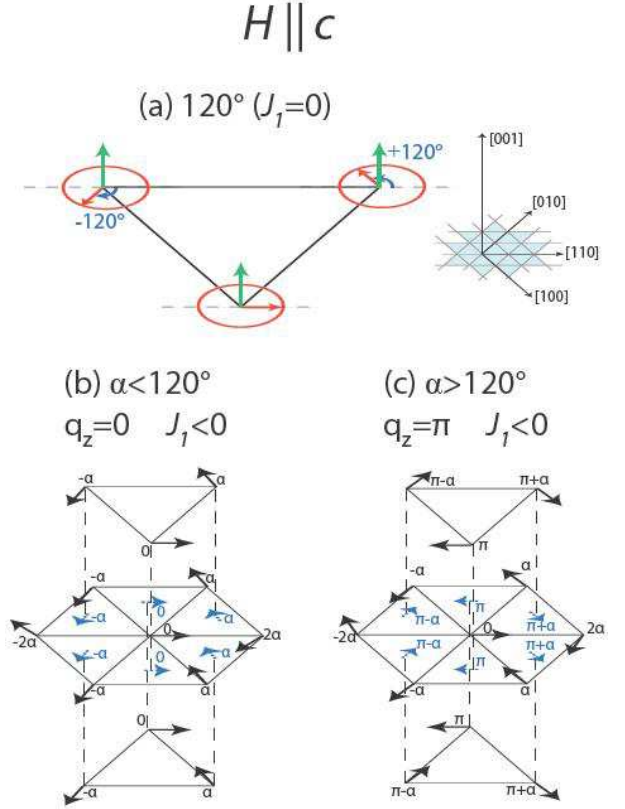}
\caption{(Color online) Schematic diagrams of the predicted spin structure for field applied along $c$-axis. (a) $120^{\circ}$ structure on a triangular plaquette corresponding to the classical solution for Heisenberg spins on a 2D lattice. Green and red arrows indicate $s^z$ and $s^{xy}$ components of the pseudospin representing each dimer unit, respectively.  Inset shows crystal axes.  Full 3D structure of $s^{xy}$ components of pseudospins on successive layers for values of $\alpha$ less than and more than $120^{\circ}$ leading to ordering wavevectors along the z-direction of 0 and $\frac{1}{2}c^*$ are plotted in (b) and (c) respectively.  Dashed blue arrows indicate the total pseudospin
moment on each triangular plaquette, illustrating ferromagnetic interplane coupling.}
\label{HParaC}
\end{figure}

Considering first the 2D lattice (effectively setting $J_1 = 0$), and noting that the $D$ anisotropy does not act on the $s^{xy}$ components of the pseudospins (i.e. $a(z)=0$), the system consists of independent triangular layers of vertical dimers, and the effective Hamiltonian reduces to
\begin{eqnarray}
\tilde{\mathcal{H}} & = & \sum_{l \langle i, j \rangle} \left[ \frac{8\left(J_2-J_3\right)}{3} \left(s_{il}^x s_{jl}^x + s_{il}^y s_{jl}^y \right) + \frac{J_2 + J_3}{2} s_{il}^z s_{jl}^z \right]
\nonumber \\
&-&  \left (g_{c c} \mu_B  H - J_0 - 3\left(J_2+J_3\right)/2\right) \sum_{l, i} s_{il}^z
\end{eqnarray}
The effective exchange anisotropy is easy-plane, i.e., the XY component of the exchange dominates. At $T=0$, the triplets condense (canted XY antiferromagnetic ordering) for $H > H_{c1} = \left(J_0-4\left(J_2-J_3\right)\right)/\left(g_{cc} \mu_B\right)$ into a state that can be approximated by a direct product of single dimer states of the form:
\begin{equation}
|\psi_{il}\rangle = \cos{\theta_{il}} |00\rangle + \sin{\theta_{il}} e^{-i\phi_{il}} |11\rangle.
\end{equation}
The canting angle $\theta_{il}=\theta$ is uniform and is set by the magnetic field, while $\phi_{il}=\phi_{l}+{\bf Q}\cdot {\bf r}_i$ with ${\bf Q}=(\pm \frac{1}{3},\pm \frac{1}{3})a^*$. The relative phase between different layers is determined by $\phi_{l}$, which can take any value for the moment because we are assuming that $J_1=0$. The expectation values of the pseudospins take a simple form
\begin{eqnarray}
\left\langle \psi_{il} | s_x |\psi_{il} \right\rangle & = &  \frac{1}{2} \sin {2 \theta} \cos {({\bf{Q}} \cdot {\bf {r_i}}+\phi_l)}
\nonumber\\
\left\langle \psi_{il} | s_y |\psi_{il} \right\rangle & = &  \frac{1}{2} \sin{2\theta} \sin{({\bf{Q}} \cdot {\bf {r_i}}+ \phi_l)}
\nonumber\\
\left\langle \psi_{il} | s_z |\psi_{il} \right\rangle & = &  -\frac{1}{2} \cos{2\theta}.
\label{ParaExpectations}
\end{eqnarray}
This corresponds to a canted antiferromagnetic state, in which the $s^{xy}$ component of the pseudospins orient  120$^{\circ}$ with respect to each other to minimize the interdimer exchange energy (i.e. the $s^{xy}$ components of the pseudospins on each triangular plaquet sum to zero) analogous to the classical solution for a Heisenberg AF on a triangular lattice.  This is illustrated in fig. \ref{HParaC}(a). The phase, corresponding to the angle of the $s^{xy}$ component of the pseudospins relative to the crystal lattice, spontaneously breaks the U(1) symmetry of the effective Hamiltonian, and the ordered state can be described as a Bose-Einstein condensate. In terms of the original spins on each Mn site, the ordered state still consists of a canted AF with the $s^{xy}$ component of the spins on adjacent dimers oriented 120$^{\circ}$ with respect to each other, but with these components reversed for spins on the top and bottom of each dimer unit.

Now consider the 3D lattice with nonzero $J_1$. The Hamiltonian still maintains U(1) symmetry, and the ordered state will still correspond to a triplet condensate (consistent with the lambda anomaly observed in heat capacity measurements for this field orientation) because the XY interaction in the Hamiltonian dominates the Ising interaction. However, the system now has the possibility to gain additional energy from the interlayer exchange. The $s^{xy}$ component of the total spin on any triangular plaquette for the classical case in fig. \ref{HParaC}(a) is zero, but if the pseudospins twist around the $z$-axis to form a spiral structure in which successive spins along the [100] and [010] directions rotate by an angle $\alpha = 120^{\circ} \pm \epsilon$ in the XY plane, the system is then able to benefit from the interlayer coupling. There are two degenerate solutions that minimize the total energy. The first solution is characterized by a uniform phase along the $c$-axis: $\phi_l=0$. In contrast, the phase is staggered $\phi_l=l \pi$, in the second solution. In addition, the shift of the single-layer ordering wave-vector from ${\bf Q}=(\pm \frac{1}{3}, \pm \frac{1}{3})a^*$ to ${\bf Q}=\pm(\alpha, \alpha)\frac{2}{\sqrt{3}a}$ (note that $a^* = \frac{4\pi}{\sqrt{3}a}$ in this non-orthogonal basis) has opposite signs for the two cases:
\begin{eqnarray}
\cos{\alpha} &=& - \frac{1}{2} - \frac{J_1}{4\left(J_2-J_3\right)} \;\; {\rm for} \;\;\phi_l=0
\nonumber \\
\cos{\alpha} &=& - \frac{1}{2} + \frac{J_1}{4\left(J_2-J_3\right)} \;\; {\rm for} \;\;\phi_l=l \pi,
\label{alfa}
\end{eqnarray}
corresponding to angles $\alpha = 111^{\circ}$ and $\alpha = 129^{\circ}$ respectively for the values of $J_1$ and $\left(J_2-J_3\right)$ obtained from single crystal INS.  The second solution was previously reported in Ref. \onlinecite{Uchida_2002}. The gain in energy of this in-plane twisting due to the inter-layer interaction, $J_1$, scales linearly with $\epsilon$ (i.e. for a given phase relation between adjacent layers there is a ``right way'' and a ``wrong way'' to twist the spiral structure). In contrast, the loss in intralayer energy from breaking the perfect 120$^{\circ}$ structure scales quadratically with $\epsilon$ because $\alpha=120^{\circ}$ is the minimum energy structure for $J_1=0$. Hence, an arbitrarily small interlayer exchange $J_1$ is able to stabilize a spiral phase with an incommensurate wave vector as indicated by eq. (\ref{alfa}). The resulting phase is incommensurate, which has been confirmed by preliminary single crystal nuclear magnetic resonance measurements and neutron scattering measurements at high fields \cite{Brown_2008, Stone_2008c}.

A similar structure, but with antiferromagnetic interplane coupling, has already been proposed by Uchida \textit{et al.}  following their initial estimation of the exchange constants in Ba$_3$Mn$_2$O$_8$ \cite{Uchida_2002}.  However, two subtleties to the ordered phase were not anticipated in that earlier paper.  The first of these is that $\epsilon$ can take both positive and negative values because there are two degenerate solutions: $\phi_l=0$ or $\phi_l=l \pi$. If $\alpha < 120^{\circ}$, as illustrated in fig. \ref{HParaC}(b), then for $J_1<0$ the component of the ordering wavevector along the $c$ direction, $q_z$, is equal to 0 ($\phi_l=0$).  However, if $\alpha > 120^{\circ}$, as  illustrated in fig. \ref{HParaC}(c), then the component of the ordering wavevector along the $c$ direction, $q_z$, is equal to $\frac{1}{2}c^*)$ ($\phi_l=l \pi$), leading to a doubling of the unit cell along the $c$-axis. A second subtlety of the ordered phase is that the resulting structure contains triangular plaquettes on which pair of spins are more closely antiferromagnetically aligned along specific directions, while other pairs of spins are less perfectly  antiferromagnetically aligned along equivalent crystallographic directions.  Specifically, adjacent spins along the [110] direction have a relative angle of $2 \alpha = 2 ( 120^{\circ} + \epsilon) \equiv 120^{\circ} - 2\epsilon$, in contrast to adjacent spins along the [100] and [010] directions which have a relative angle of $\alpha = 120^{\circ} + \epsilon$. Magnetostriction associated with this ``bond ordering" must lead to a subtle lattice deformation, characterized by a separate but related order parameter. If the coupling to the lattice is strong enough, we can even anticipate that the phase transition will become weakly first order \cite{Ginzburg}.  Microscopic energetics of the two different bond ordering scenarios will break the degeneracy  and determine the $c$-axis ordering wavevector.

\subsection{Fields perpendicular to $c$}

\begin{figure}
\includegraphics[width=8cm]{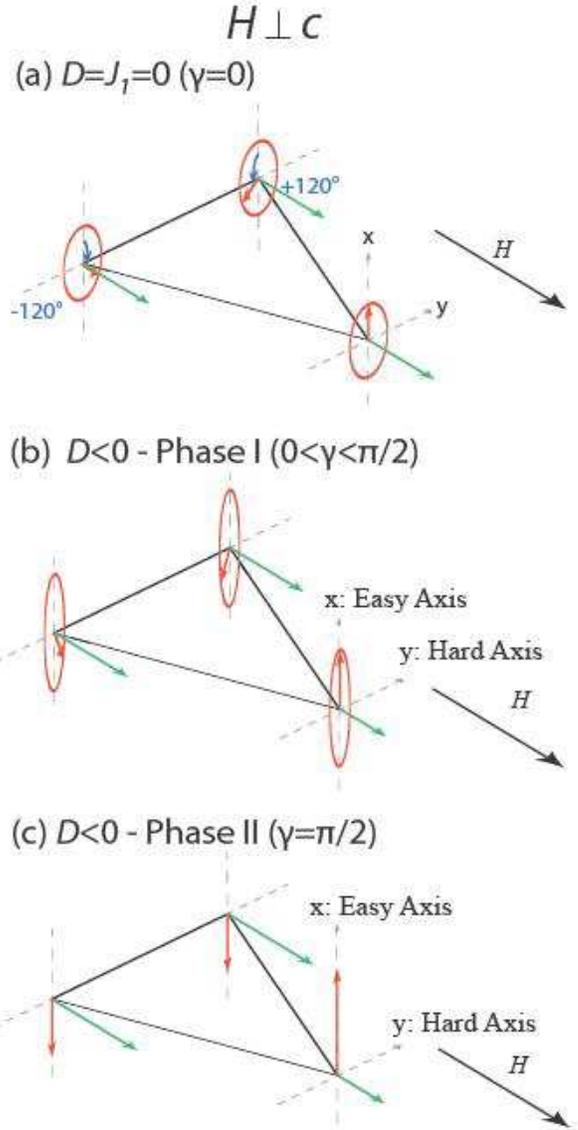}
\caption{(Color online) Schematic diagram showing spin structure for fields pointing along the `point' of a triangular plaquette obtained by minimizing parameters in eq. \ref{PerpExp}. Green and red arrows indicate $s^z$ and $s^{xy}$ components of the pseudospin representing each dimer unit, respectively.  (a) Spin structure for $D$=0 and $J_1$=0 for $H\|$[110]. The $s^{xy}$ components of the pseudospins are oriented $120^{\circ}$ from each other, equivalent to the case shown in figure \ref{HParaC}(a) for $H \| c$, but rotated into the [010]-[001] plane. (b) Partially modulated phase (Phase I) for $D<0$ and $J_1$ finite.  The $s^{xy}$ components of the spin precess along an elliptical path with an incommensurate wavevector close to $(\frac{1}{3}, \frac{1}{3})a^*$. (c) Maximally modulated phase (Phase II) stabilised close to $H_{c1}$ and $H_{c2}$.}
\label{ThreeTri}
\end{figure}

For fields oriented away form the $c$-axis, the anisotropy term $D$ breaks the U(1) symmetry of the Hamiltonian, and can stabilize an Ising-like modulated structure. To understand the nature of this phase, it is instructive to first consider the case in which $D$ vanishes on a 2D lattice ($J_1 =0$) and a magnetic field is applied perpendicular to the crystalline c-axis, for instance along the direction midway between [100] and [110] (i.e. along the ``point" of a triangular plaquette). All of the same arguments given above for the case $H \| c$ and $J_1 = 0$ still apply, but the quantization axis now lies in the ab plane so the XY order lies in the plane defined by the two vectors [001] and [010] (fig. \ref{ThreeTri}(a)). Since $D=0$, there is no anisotropy in this plane, and the pseudospins spontaneously break U(1) symmetry -- the ordered state is a Bose Einstein condensate. However, a finite value of $D$ in equation \ref{EffHam} qualitatively changes the nature of the ground state. For the specific example of the field oriented along the tip of the triangular plaquette, a negative value for $D$, appropriate for Ba$_3$Mn$_2$O$_8$, implies an easy axis for the $s^{xy}$ component of the the pseudospins along the [001] direction. For the 2D lattice (i.e. $J_1=0$), the in-plane interaction between spins still favors a magnetic structure for which the total spin in the XY plane vanishes. To minimize both the anisotropy energy and also the in-plane exchange energy, the system may adopt an inhomogeous magnetic structure in which the component of the pseudospins along the hard axis are depressed relative to along the easy axis while the canting angle along the $z$-axis is adjusted so as to preserve zero net $s^{xy}$ spin on each triangular plaquette (fig. \ref{ThreeTri}(b)).

\begin{figure}
\includegraphics[width=8cm]{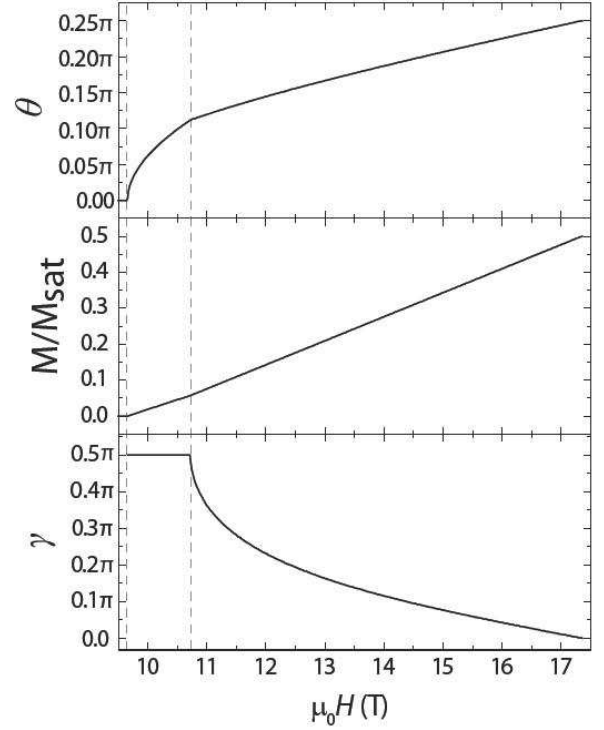}
\caption{Variational parameters $\gamma$ and $\theta$ and normalized magnetization $M/M_{sat}$ that result from minimizing the energy (${\tilde {\cal H}}$) for the spin configuration described by eq. (\ref{PerpExp}). The vertical dashed lines mark $H_{c1}$ and the transition between Phase II and Phase I.  We have used values of $J_0=19.2$ K, $J_1=-1.39$ K and $J_2-J_3=1.26$ K that result from preliminary fits of triplet dispersion measured at $H=0$ in a single crystal of Ba$_3$Mn$_2$O$_8$ \cite{Stone_2008b}. The value of $J_3=1.27$ K was chosen to fit the measured optimal field: $\left(H_{c1}+H_{c2}\right)/2 \sim$ 17.3 T.  The disagreement between the calculated $H_{c1} \sim 9.6$ T and the measured value of $\sim 8.7$ T is due to our two-level (singlet-triplet) approximation. Good quantitative agreement is obtained when the other two triplets are included \cite{Batista_2008}.}
\label{variational}
\end{figure}

Considering now the 3D lattice (finite $J_1$), the general form of the pseudospins describing this modulated behavior  in a spin dimer system with uniaxial anisotropy is
\begin{eqnarray}
\left\langle \psi_{il} | s_x |\psi_{il} \right\rangle & = &  \frac{1}{2} \sin {2 \theta} \cos {({\bf{Q}} \cdot {\bf {r_i}}+\phi_l)}
\nonumber\\
\left\langle \psi_{il} | s_y |\psi_{il} \right\rangle & = &  \frac{1}{2} \cos{\gamma}\sin{2\theta} \sin{({\bf{Q}} \cdot {\bf {r_i}}+\phi_l )}
\nonumber\\
\left\langle \psi_{il} | s_z |\psi_{il} \right\rangle & = &
\frac{\pm 1}{2}\sqrt{\cos^2{2\theta}+\sin^2{2\theta} \sin^2{\gamma} \sin^2{({\bf{Q}} \cdot {\bf {r_i}}+ \phi_l )}},
\nonumber\\
\label{PerpExp}
\end{eqnarray}
where $0 \leq \gamma \leq \pi/2$ sets the ratio between the maximum amplitude of the $s^y$ and $s^x$ components and consequently the amplitude of the modulation of the $s^z$ component ($\gamma=0$ is unmodulated and $\gamma = \pi/2$ is maximally modulated).  The positive (negtive) sign in the last line of eq. \ref{PerpExp} holds for $\theta > \pi/4$ ($\theta<\pi/4$). Minimization of the Hamiltonian (\ref{EffHam}) with respect to the different parameters $\theta$, $\gamma$, $\bf{Q}$ and $\phi_l$ yields a ground state. The optimal values of ${\bf Q}$ and $\phi_l$ are still very well approximated by eq. (\ref{alfa}). The optimal values of $\theta$ and $\gamma$ as a function of the field $H$ are shown in fig.\ref{variational}. The ordered ground state has no $s^y$ component of the pseudospin ($\gamma=\pi/2$) for $H$ near $H_{c1}$ (fig. \ref{ThreeTri}(c)). However the effective exchange anisotropy in $\tilde{\mathcal{H}}$ (eq. \ref{EffHam}) penalizes the modulation of the $s^z$ component and favors a less modulated structure ($\gamma<\pi/2$) when the $z$-component of the real magnetization becomes large enough, i.e., when $H-H_{c1}$ is large enough.  This is presumably the origin of the two distinct phases observed in thermodynamic measurements for fields oriented perpendicular to the $c$-axis, and indeed a full analysis including all three triplet states quantitatively accounts for $H_{c1}$ \cite{Batista_2008}. Even without including these terms the agreement with the measured phase diagram is remarkable.  According to our results shown in  fig. \ref{variational}, the transition between both phases is of second order at $T=0$. For $H$ slightly larger than $H_{1,2}$ (critical field for the transition between phases I and II), we obtain the field dependence $\pi/2-\gamma \propto \sqrt{H-H_{1,2}}$ characteristic of a mean field transition. Correspondingly, the total magnetization and $\theta$ exhibit a kink at $H_{1,2}$. The resulting structure stable at higher fields is still modulated along the $z$ direction but with a finite component along the $y$ direction (fig. \ref{ThreeTri}(b)).  The $s^y$ component of the pseudospin varies to the $s^x$ component as a function of field, yielding an unmodulated structure exactly at the middle of the dome since $s^z$=0 at this field, equivalent to the $H \| c$ structure.  At this field, rotation of field into the $H \| c$ direction therefore occurs without crossing a phase boundary, consistent with our labelling of Phase I in fig. \ref{PhaseB}(a) and (b).  The energy associated with the anisotropy $a(x) \left(J_2-J_3\right) \sim D \left(J_2-J_3\right) / J_0 \sim $25mK  is small, consistent with the observation of a $\lambda$-like anomaly in heat capacity seen in fig. \ref{Cp}(b) (i.e the critical scaling associated with the Ising phase transition will only apparent very close to $T_c$).

The resulting modulated structures are characterized by three separate order parameters: there is a finite modulation of the magnetization along the field direction (Ising order), there is also Ising ordering along the easy direction perpendicular to the field, and finally there is bond order. Referring back to the heat capacity data (fig. \ref{Cp}), the anomaly associated with this phase transition into Phase II is clearly different to the $\lambda$ anomaly seen for fields oriented parallel to the $c$-axis, although the data do not permit a critical scaling analysis. Estimates of the change in entropy through the transition into Phase II vary only weakly with field or temperature, implying a significant lattice contribution.

\subsection{Interplane frustration: Comparison to BaCuSi$_2$O$_6$}

\begin{figure}
\includegraphics[width=8cm]{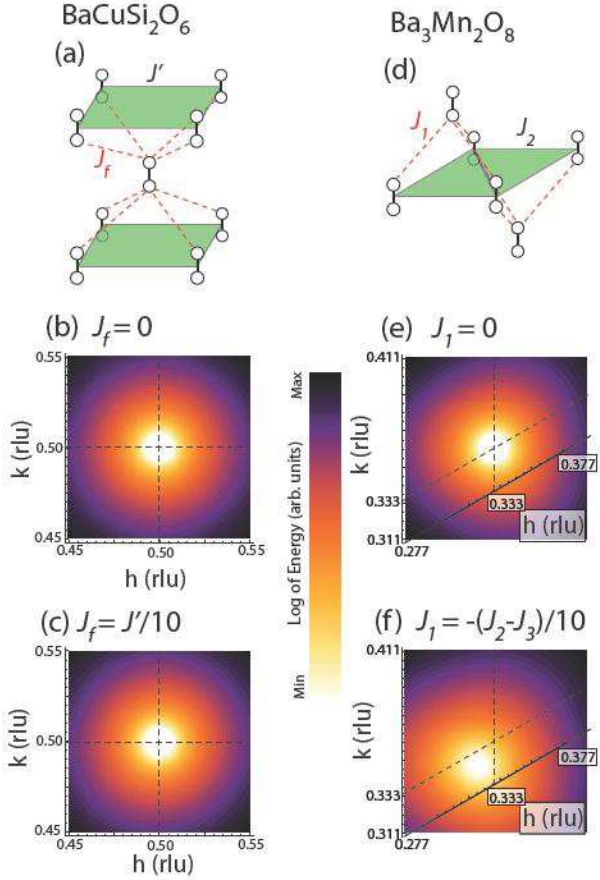}
\caption{(Color online) Comparison of the effect of interlayer coupling on the dispersion minimum for BaCuSi$_2$O$_6$ (panels a-c) and Ba$_3$Mn$_2$O$_8$ (panels d-f).  (a) Idealized body centered tetragonal lattice of spin dimer compound BaCuSi$_2$O$_6$ showing the intralayer exchange $J'$ and the interlayer exchange $J_f$. (b) and (c) show BaCuSi$_2$O$_6$ triplet dispersion minimum at (1/2, 1/2) for interlayer exchange $J_f$ equal to 0 and 0.1$J'$, respectively.  (d) Triangular lattice of Ba$_3$Mn$_2$O$_8$ showing intralayer exchanges $J_2$ and interlayer exchange $J_1$.  $J_3$ not shown for clarity (see fig. \ref{XtalSuscept}(a)). (e) Ba$_3$Mn$_2$O$_8$ triplet dispersion for the case $J_1$=0, showing the minimum located at (1/3,1/3). (f) Including a small interlayer coupling, shown here for the specific case of $J_1=-(J_2-J_3)/10$, shifts the minimimum of the dispersion away from (1/3,1/3).}
\label{Frust}
\end{figure}

It is instructive to compare the proposed magnetic structures of Ba$_3$Mn$_2$O$_8$ with those of the closely related spin dimer compound BaCuSi$_2$O$_6$. Both materials have a small frustrated interlayer coupling \cite{Ruegg_2007, Stone_2008}. In the case of BaCuSi$_2$O$_6$, vertical dimers are arranged on a (nominally \cite{Samulon_2006}) body centered tetragonal (bct) lattice \cite{Jaime_2004}. The dispersion of the triplet associated with a single square layer of dimers, derived from the effective Hamiltonian for this system (see eq.(2) in ref. \onlinecite{Jaime_2004}), has planar inversion symmetry about the 2D minimum, which is located at $\bf{Q}=(\frac{1}{2},\frac{1}{2})a^*$ (fig. \ref{Frust}a)). Therefore, an expansion of the single particle dispersion about the point $\bf{Q}$ including the weaker interlayer coupling only contains terms quadratic in \textbf{k}:
\begin{equation}
E\left(\bf{Q}+\textbf{k}\right) - E\left(\bf{Q}\right) \sim  J'(k_x^2 + k_y^2) + 2J_fk_xk_y
\end{equation}
where we have used the same labeling of the in-plane interdimer coupling ($J'$) and the inter-plane coupling ($J_f$) introduced in ref \cite{Sebastian_2006a}. Significantly, since $J_f$ is an order of magnitude smaller than $J'$, the presence of finite interlayer exchange does not move the minimum of the dispersion from $\bf{Q}=(\frac{1}{2},\frac{1}{2})a^*$ (fig. \ref{Frust}(b)). The resulting interlayer frustration of the perfect bct lattice is partially lifted by the effect of quantum fluctuations \cite{Maltseva_2005}, resulting in a fully three dimensional magnetic structure away from the QCP. However, since the quantum phase transition is driven by amplitude fluctuations, the QCP itself is two-dimensional \cite{Sebastian_2006a, Batista_2007}. Additional terms in the Hamiltonian describing the real material (including deviations from the perfect bct lattice \cite{Samulon_2006}, and the effects of the higher energy triplet states \cite{Rosch_2007}) will partially lift the perfect frustration even at $T = 0$. Nevertheless, the observation in this material of 2D critical scaling exponents close to $H_{c1}$ \cite{Sebastian_2006a} indicates that there is a considerable region over which the fluctuations are determined by the 2D fixed point.

The model describing Ba$_3$Mn$_2$O$_8$ has markedly different properties. In this case, the minimum of the dispersion relation associated with a single triangular layer, derived from the effective Hamiltonian (eqn. \ref{EffHam}), is located at $\bf{Q}=(\frac{1}{3},\frac{1}{3})a^*$ referred to the primitive unit vectors of the lattice \cite{Stone_2008}. An expansion of the single particle dispersion about the point $\bf{Q}$ now contains terms linear in $\textbf{k}$:
\begin{eqnarray}
E\left(\bf{Q}+\textbf{k}\right) - E\left(\bf{Q}\right) & \sim & \left(J_2-J_3\right) \left(k_x^2 + k_y^2+k_xk_y\right)
\nonumber\\
&& + \frac{\sqrt{3}}{2}J_1\left(k_x+k_y\right).
\end{eqnarray}
The competing effects of the linear and quadratic terms means that arbitrarily small interplane coupling $J_1$ shifts the minimum of the dispersion from $\bf{Q}=(\frac{1}{3},\frac{1}{3})a^*$, illustrated in fig. \ref{Frust}(e,f). This has the dual effect of establishing an incommensurate  spiral structure at low temperatures, and lifting the perfect interlayer frustration of the $120^{\circ}$ structure. In effect, the triangular lattice is less frustrated than the bct lattice, at least with regard to the effect of interlayer coupling! Hence, for Ba$_3$Mn$_2$O$_8$ we anticipate a fully three dimensional magnetic structure down to $T=0$, and three dimensional critical scaling exponents. Experiments to determine these are in progress.

\section{\label{sec:level5}Conclusion}

In summary, via a combination of heat capacity, MCE and cantilever torque magnetometry measurements, we have established the low temperature phase diagram of the $S=1$ spin dimer compound Ba$_3$Mn$_2$O$_8$ which is associated with closing of the spin gap to excited triplet states. These data reveal two distinctly different ordered states (labelled as Phases I and II) for fields oriented perpendicular to the crystalline $c$-axis. For fields oriented parallel to $c$, only Phase I is observed, at least for temperatures above 300mK. Analysis of the data, and consideration of the minimal spin Hamiltonian that describes the system, indicates that Phase I consists of a spiral structure which is stabilized by the weak interlayer coupling. For fields perpendicular to $c$ this phase is partially modulated due to the single ion anisotropy.  In contrast, Phase II appears to be a fully modulated structure with no  moment along the $y$ direction, stabilized by a combination of the single ion anisotropy and the interlayer coupling. Both phases implicitly contain spontaneous bond ordering, and are characterized by multiple independent order parameters. Experiments are in progress to directly determine the magnetic structures of which we described the qualitative features.

\section{Acknowledgments}

The authors thank S. E. Brown, S. Hill, M. D. Lumsden and M. B. Stone for useful discussions.  Work at Stanford University is supported by the National Science Foundation, Division of Materials Research under grant DMR-0705087. Crystal growth equipment purchased with support from the Department of Energy, Office of Basic Energy Sciences, under contract DE-AC02-76SF00515.  Experiments performed at the NHMFL were supported by the NSF, the State of Florida, and the Department of Energy.

\appendix

\section{Derivation of the effective exchange anisotropy}

Here we derive the effective exchange anisotropy term of ${\tilde {\cal H}}$ that is induced by the single-ion anisotropy term $H_{SI}= D \sum_{i,\mu}  (S^x_{i \mu})^2$ of ${\cal H}$. There are two different processes that contribute to the amplitude, $2a(x)\left(J_2-J_3\right)$, of the effective exchange anisotropy. In the first process, $H_{SI}$ induces an intradimer transition between the singlet $|00\rangle$ and the quintuplet $|22\rangle$ and the corresponding matrix element is:
\begin{equation}
\langle 22| D [(S^x_{1})^2 + (S^x_{2})^2] |00\rangle = \frac{D}{\sqrt{3}}.
\end{equation}
The gap between the singlet and $|22\rangle$, for a field $H$ such that the ($|11\rangle$) triplet and the singlet states are degenerate,  is $J_0$. The second step is a transition between the state $|22\rangle_i |00\rangle_j$ on nearest-neighbor dimers $i,j$ and the state with two $S^z_1+S^z_2=1$ triplets $|11\rangle_i |11\rangle_j$ produced by the interdimer Heisenberg interactions $J_2$ and $J_3$. The corresponding matrix element is:
\begin{equation}
\langle 11|_j \langle 11 |_i  H_{J_2} + H_{J_2}  |22\rangle_i |00\rangle_j = \frac{2\left(J_2-J_3 \right)}{\sqrt{3}}
\end{equation}
There is a factor of 2 that results from the fact that the quintuplet state can be created in any of the two dimers ($i$ or $j$) involved in this second order process.

The second contribution to $2a(x)\left(J_2-J_3\right)$ comes from the following second order process: the pair of singlet states can be excited into a pair of triplets with opposite $S^z_1+S^z_2$,
\begin{equation}
\langle 11|_j \langle 1{\bar 1} |_i  H_{J_2} +H_{J_3} |00\rangle_i |00\rangle_j = \frac{4\left(J_2-J_3\right)}{3},
\end{equation}
and the triplet state with $S^z_1+S^z_2=-1$ can be flipped into the $S^z_1+S^z_2=1$ by the $D$ term,
\begin{equation}
\langle 1{\bar 1}| D [(S^x_{1})^2 + (S^x_{2})^2] |11\rangle = \frac{D}{2}.
\end{equation}
Again there are two of these processes because the $S^z_1+S^z_2=-1$ triplet can be in any of both dimers.

The sum of these contributions leads to an effective exchange anisotropy:
\begin{equation}
2 \left(J_2-J_3\right) a(x) \sum_{l,\langle i,j \rangle} (s^x_{il} s^x_{jl}-s^y_{il} s^y_{jl})
\end{equation}
with
\begin{equation}
a(x)= - \frac{8D}{3J_0}.
\end{equation}
By following a similar procedure, we obtain an effective exchange anistropy for the inter-layer coupling:
\begin{equation}
J_1 a(x) \sum_{l,\langle\langle i,j \rangle\rangle} (s^x_{il} s^x_{jl}-s^y_{il} s^y_{jl}).
\end{equation}
These are the two anisotropic terms that appear in the effective Hamiltonian ${\tilde {\cal H}}$ of eq.(\ref{EffHam}).


\begin{thebibliography}{1}


\bibitem{Anderson_1973} P. W. Anderson, Mat. Res. Bull. \textbf{8}, 153 (1973).
\bibitem{Jolicoeur_1989} T. Jolicoeur and J. C. Leguillou, Phys. Rev. B \textbf{40}, 2727 (1989).
\bibitem{Bernu_1994} B. Bernu, P. Lecheminant, C. Lhullier, and L. Pierre, Phys. Rev. B \textbf{50}, 10048 (1994).
\bibitem{Diep_2005} See for example \textit{Frustrated Spin Systems} Ed H. T. Diep and refs therein.
\bibitem{Sachdev_2004} S. Sachdev in \textit{Quantum Phase Transitions} (Springer, Berlin, 2004).
\bibitem{Matsubara_1956} T. Matsubara and H. Matsuda, Prog. Theor. Phys. \textbf{16}, 569 (1956).
\bibitem{Rice_2002} T. M. Rice, Science \textbf{298}, 760 (2002).
\bibitem{Sebastian_2006a} S. E. Sebastian, N. Harrison, C. D. Batista, L. Balicas,  M. Jaime, P. A. Sharma, N. Kawashima, and I. R. Fisher, Nature \textbf{411} 617 (2006).
\bibitem{Batista_2007} C. D. Batista, J. Schmalian, N. Kawashima, P. Sengupta, S. E. Sebastian, N. Harrison, M. Jaime, and I. R. Fisher, Phys. Rev. Lett. \textbf{98} 257201 (2007);
J. Schmalian and C. D. Batista, Phys. Rev. B {\bf 77}, 094406 (2008).
\bibitem{Sebastian_2006b} S. E. Sebastian, P. Tanedo, P. A. Goddard, S.-C. Lee, A. Wilson, S. Kim, S. Cox, R. D. McDonald, S. Hill, N. Harrison, C. D. Batista, and I. R. Fisher, Phys. Rev. B \textbf{74} 180401(R) (2006).
\bibitem{Sengupta_2007} P. Sengupta and C. D. Batista, Phys. Rev. Lett. \textbf{98}, 227201 (2007).
\bibitem{Weller_1999} M. T. Weller and S. J. Skinner, Acta Crystallogr., Sect. C: Cryst. Struct. Commun. \textbf{55}, 154 (1999).
\bibitem{Uchida_2002} M. Uchida, H. Tanaka, H. Mitamura, F. Ishikawa, and T. Goto, Phys. Rev. B \textbf{66}, 054429 (2002).
\bibitem{Stone_2008} M. B. Stone, M. D. Lumsden, Y. Qiu, E. C. Samulon, C. D. Batista, and I. R. Fisher, 	arXiv:0801.2332v2.
\bibitem{Stone_2008b} M. B. Stone {\it et al}, unpublished.
\bibitem{Interaction} Although the out-of-plane exchange $J_1$ is slightly larger than the in-plane interdimer exchange $J_2$, the in plane hopping dominates because there are twice as many nearest neighbors.
\bibitem{Whitmore_1993} M. H. Whitmore, H. R. Verd\'un, and D. J. Singel, Phys. Rev. B \textbf{47}, 11479 (1993).
\bibitem{Hill_2007} S. Hill, private communication.
\bibitem{Tsujii_2005} H. Tsujii, B. Andraka, M. Uchida, H. Tanaka, and Y. Takano, Phys. Rev. B \textbf{72}, 214434 (2005).
\bibitem{Sebastian_2005} S. E. Sebastian, P. A. Sharma, M. Jaime, N. Harrison, V. Correa, L. Balicas, N. Kawashima, C. D. Batista, and I. R. Fisher, Phys. Rev B \textbf{72}, 100404(R) (2005).
\bibitem{Universality} The critical exponents $\alpha$ associated with divergent behavior of the heat capacity is very small for both the 3DXY ($\alpha=0.01$) and Ising ($\alpha=0.1$) universality classes. A single relevant exponent will descrive the divergent behavior irrespective of which direction the phase boundary is crossed (i.e. varying field or temperature), in which case a peak would be expected in the third derivative of the free energy, corresponding to the second derivative of the magnetization.
\bibitem{Brown_2008} S. E. Brown, private communication.
\bibitem{Stone_2008c} M. B. Stone and M. D. Lumsden, private communication.
\bibitem{Ginzburg} In a Ginzburg-Landau free energy expansion, the lattice distortion adds an additional negative term quartic in the order parameter. If this term is large enough to outweigh the original positive quartic term, the system suffers a first order transition on cooling.
\bibitem{Batista_2008} C. D. Batista and P. Sengupta, unpublished.
\bibitem{Ruegg_2007} Ch. R\"uegg, D. F. McMorrow, B. Normand, H. M. R\o nnow, S. E. Sebastian, I. R. Fisher, C. D. Batista, S. N. Gvasaliya, Ch. Niedermayer, and J. Stahn, Phys. Rev. Lett. \textbf{98} 017202 (2007).
\bibitem{Samulon_2006} E. C. Samulon, Z. Islam, S. E. Sebastian, P. B. Brooks, M. K. McCourt Jr., J. Ilavsky, and I. R. Fisher, Phys. Rev. B \textbf{73} 100407(R) (2006).
\bibitem{Jaime_2004}M. Jaime, V. F. Correa, N. Harrison, C. D. Batista, N. Kawashima, Y. Kazuma, G. A. Jorge, R. Stern, I. Heinmaa, S. A. Zvyagin, Y. Sasago, and K. Uchinokura, Phys. Rev. Lett. \textbf{93}, 087203 (2004).
\bibitem{Maltseva_2005} M. Maltseva and P. Coleman, Phys. Rev. B \textbf{72}, 174415(R) (2005).
\bibitem{Rosch_2007} O. R\"osch and M. Vojta, Phys. Rev. B \textbf{76}, 180401(R) (2007).

\end{thebibliography}
\end{document}